\documentstyle[sprocl]{article}

\def\Journal#1#2#3#4{{#1} {\bf #2}, #3 (#4)}

\def\PLB{{\em Phys. Lett.}  B}

\def\be{\begin{equation}}
\def\ee{\end{equation}}
\def\bea{\begin{eqnarray}}
\def\eea{\end{eqnarray}}

\begin{document}

\hspace{8.cm}{nucl-th/9709031}

\hspace{8.cm}{La Plata-Th 97/21}

\title{THE MIT BAG MODEL IN NUCLEAR MEDIUM FROM EQMC AND QHD DESCRIPTIONS
\footnote{To be published in the Proceedings of the
V Workshop on Relativistic Aspects of
Nuclear Physics, CBPF, Brazil, 1997.}}

\author{ MARTIN SCHVELLINGER }

\address{Department of Physics, Universidad Nacional de La Plata,
C.C. 67 (1900), \\ La Plata, Argentina}

\maketitle\abstracts{
Recent developments in the density dependence of the MIT bag radius and
parameters in nuclear matter are discussed. Starting from the MIT bag
lagrangian density, the calculations are specialized for symmetric
homogeneous isotropic nuclear matter. A simultaneous description
of the density dependence of MIT bag model is derived from the Quantum
Hadrodynamics and an extended Quark Meson Coupling model.
Results are in agreement with those derived from Scaling Model.}

\section{Introduction}

As it is well known, the Quantum Chromodynamics (QCD) is the
most accepted candidate for a theory of strong interactions.
Although QCD has been successful in describing the high energy
domain, its application seems difficult in the energy range
associated to nuclear phenomena.
Several QCD-inspired models have been developed to sort up
the non-perturbative character of QCD at the nuclear energy range.
For instance, there are the so-called bag models. Although these
kind of models are made up to describe hadron properties, in general
they have a set of parameters which must be fixed in order to reproduce
certain specific hadron or nuclear properties.
In order to describe hadrons in nuclear medium it is very interesting
to known the density dependence of the parameters of the specific model.
In this presentation we want to do that for the MIT bag model.
In this sence, we have used an extended version of the Quark Meson
Coupling model (EQMC).
In the present case we assume the validity of the nuclear matter
predictions derived from an effective hadronic lagrangian,
looking for the implications that this description has for a
picture with deals with subnuclear degrees of freedom, like that EQMC.

\section{Formalism}

The MIT bag lagrangian density is
\begin{eqnarray*}
{\cal{L}}^{MIT}(x)= \sum_{\alpha=1}^3 [ {\bar{q}}_{\alpha}(x)
(i {\gamma}_{\mu}{\partial}^{\mu} - m_q + g^q_{\sigma} {\sigma}(x) \\
- g^q_{\omega} {\gamma}_{\mu} {\omega}^{\mu}(x)) q_{\alpha}(x) -
\frac{B}{3}] {\Theta}_V - \sum_{\alpha=1}^3 \frac{1}{2}
{\bar{q}}_{\alpha}(x) q_{\alpha}(x) {\Delta}_S \\
+ \frac{1}{2} ({\partial}^{\mu}{\sigma}(x){\partial}_{\mu}{\sigma}(x)
-m^2_{\sigma} {\sigma}^2) + \frac{1}{2} m^2_{\omega} {\omega}_{\mu}(x)
{\omega}_{\mu}(x) - \frac{1}{4} F^{\mu \nu}(x) F_{\mu \nu}(x) ,
\end{eqnarray*}
where the degrees of freedom are the quark fields
$q_{\alpha}(x)$ (up or down), the scalar-isoscalar meson field
$\sigma (x)$ and the vector-isoscalar field ${\omega}_{\mu} (x)$.
$F^{\mu \nu}(x)$ is defined as
${\partial}^{\mu} {\omega}^{\nu}(x)
- {\partial}^{\nu} {\omega}^{\mu}(x)$.
The bag boundary conditions are used and the
Euler-Lagrange eqs. for quarks, and mesons are derived.
$g^q_{\sigma}$ and $g^q_{\omega}$ are the quark-meson coupling
constants. The continuity of the fields through the bag surface
is assumed. The Mean Field approximation is used, and in this form we
have the MIT bag energy
\begin{eqnarray*}
E_b = \int d^3x  T^{00}_{MFA} = \frac{4}{3} \pi R^3 B + \nonumber \\
\frac{i}{2} \sum_{\alpha=1}^3 \int_V d^3x
<{\bar{q}}^{\dagger}_\alpha \partial^0 q_\alpha -
\partial^0 {\bar{q}}^{\dagger}_\alpha q_\alpha>
+ \frac{2}{3} \pi R^3 (m_\sigma {\bar{\sigma}}^2 -
m_\omega {\bar{\omega}}^2),
\end{eqnarray*}
where $T^{\mu \nu}$ is the energy momentum tensor,
$q_\alpha({\vec{r}},t)$ is
the normalized ground-state quark wave function for a spherical bag
of radius $R$ expressed in terms of the $y$-variable calculated from
the boundary conditions for the bag, the bag radius $R$,
the effective quark mass given by
$m^*_q = m_q -g^q_{\sigma} {\bar{\sigma}}$ and the
effective energy eigenvalue
${\epsilon}_q = \Omega/R+ g^q_{\omega} {\bar{\omega}}$,
where ${\Omega} = \sqrt{y^2 + (R m^*_q)^2}$.

The bag energy is
$E_b = {{3 {\Omega} - z_0}\over{R}} + {{4}\over{3}} \pi B R^3$
where $B$ is the energy per unit of volume and $z_0$ takes into
account the zero point energy of the bag.
The nucleon mass (including the correction due to the
spurious center of mass motion) is $M^*_b = \sqrt{E_b^2 - 3 {(y/R)}^2}$.
Here $M^*_N$ is the nucleon mass entering in QHD and
$M^*_b$ is the nucleon mass generated by the bag model $M^*_b$.
The pressure inside the bag is $P_b=- \partial E_b / \partial V$.

Always the bag parameters at zero baryon density are calculated
to reproduce the experimental nucleon mass $M_b = 939$ MeV.
Simultaneously it is required the equilibrium condition for the bag
$d M_b( \sigma )/d R = 0$, evaluated at $\sigma=\bar{\sigma}$.
By using the bag radius at zero baryon density $R_0$, and the quark
mass $m_q$ as free parameters, one can obtain the values shown in
reference \cite{SAITO1}.
In the present case, since the bag is immersed in the nuclear
medium, the bag is not isolated, we need calculate the nucleon mass
and the bag pressure extracted from the nuclear medium
\cite{EQMC1,EQMC2}, i.e., using the Quantum Hadrodynamics (QHD)
\cite{WALECKA}.
In QHD, nucleons $\psi$ and mesons $\sigma, \omega_{\mu}, \pi, b_{\mu}$
are the relevant degrees of freedom. In the simplest version
(the so-called QHD-I model) nucleons and only scalar ($\sigma$)
and vector ($\omega_{\mu}$) neutral mesons are used. We have used
the lagrangian density corresponding to the Zimanyi-Moszkowski
model (ZM) \cite{ZM}.
We have used the MFA and calculated
the effective nucleon mass $M^*_N = M_N - V({\bar{\sigma}})$ and
the hadronic pressure for uniform nuclear matter
$P_h = - {{1}\over{3}} T^{ii}$ (where $T^{ii}$ is the trace over
the spatial components of the energy-momentum tensor $T^{\mu \nu}$).

If EQMC and QHD produce coherent descriptions, the following
condition must be fulfilled $M^*_N( \sigma ) = M^*_b ( \sigma )$,
together with $g_{\sigma}=3 g^q_{\sigma}$, $g_{\omega}=3 g^q_{\omega}$.
The stability of the bags
in the nuclear medium with respect to volume changes
is impossed by $P_b( \sigma )=P_h( \sigma )$,
where $P_b( \sigma )$ is the internal pressure generated by the
quark dynamics and $P_h( \sigma )$ is the external hadronic pressure.
The last equation is a statistical equilibrium condition
on the bag surface which ensures a direct relation between nuclear
matter bulk properties and the stability of the confining volume.
These relations have been used to get the density variation
of the bag parameters. They can be expanded in power series and
by equating coefficient to coefficient additional relations can
be obtained. Since in our approach only the linear contributions
have been retained, we have additional equations
$\left. {{d M^*_b(\sigma)}\over{d \sigma}} \right |_{\bar{\sigma}} = -
\left. {{d V(\sigma)}\over{d \sigma}}\right |_{\bar{\sigma}}$
and
$\left. {{dP_b(\sigma)}\over{d \sigma}}\right |_{\bar{\sigma}} =
\left. {{dP_h(\sigma)}\over{d \sigma}}\right |_{\bar{\sigma}} $
for each ${\bar{\sigma}}$ at a given density.
Since the $\sigma$-dependence is contained in the bag parameters
$R$, $B$ and $z_0$, the hypotesis is equivalent to assume that the
derivatives $\lambda=dB/d\bar{\sigma}$ and
$\mu=dz_0/d\bar{\sigma}$ are constant values.

\section{Results and conclusions}

To search for appropriate values of $\lambda$ and $\mu$
we have explored the $(\lambda, \mu)$ plane at zero
baryon density. $R, B$ and $z_0$ at zero baryon
density as a function of $\lambda$ for several values of $\mu$
and using the ZM model of QHD.
We have selected two sets of values
set I ($\lambda=-5.28$ fm$^{-3}$,
$\mu=-0.50$ fm); set II ($\lambda=0, \mu=1.6$ fm)
which places the ZM model in the regions I and II, respectively.
The bag radius as a function of the density has been calculated.
We have compared our results with those obtained by Jin and Jennings
\cite{XJ1}. In ref. \cite{XJ1} the density dependence of the bag constant
has been modeled in two different forms: the so-called
Direct Coupling Model (DCM) and the Scaling Model (SM).
The bag constant is parametrized by
$\frac{B}{B_0} = {\left[ 1 - 4 \frac{g^B_{\sigma}
{\bar{\sigma}}}{\delta M_N} \right]}^{\delta}$, for the DCM and
$\frac{B}{B_0} = {\left[ \frac{M^*_N}{M_N}
\right]}^{k}$ for the SM. $g_{\sigma}^B$, $\delta$ and $k$ are positive
parameters and $B_0$ is the bag parameter at zero density.
The table shows a comparison between our results by using ZM model
with the set II and the ones of ref. \cite{XJ1} with the DCM.
The last row corresponds to our calculations
with ZM model and set II, the remaining rows display the DCM
results corresponding to the model parameters indicated.
This density dependece was fitted with a quadratic polynomial
$B=A_0+A_1 {\bar{\sigma}}+A_2 {\bar{\sigma}}^2$.
The best fit has been obtained for $A_0=184,723$ MeV,
$A_1=-186.42$ MeV and $A_2=112.37$ MeV.
These values correspond to
$g_{\sigma}^q=4.8$ and $\delta = 20.5$ for the
DCM in the same degree of approximation.
The best fit of our results is for the value of
$k=3.16$ for the SM.
Our results are in good agreement with the SM model and
they coincide with the DCM model only
for small values of ${\bar{\sigma}}$.

To conclude, I want to remark that
we have studied the coherence of the QHD and QMC descriptions
by using the equilibrium conditions for the MIT bag in
nuclear matter.
The density dependence of the bag parameters
and the bag radius have been evaluated by using two dynamical
quantities,
i.e. the derivatives $dB/d \bar{\sigma}$ and $dz_0/d \bar{\sigma}$,
as parameters and we have explored their possible variation range
\cite{EQMC1,EQMC2}.
We have found two different dynamical regimes for these parameters.
We have chosen the Zimanyi-Moszkowski model to study the EMC effect
because it is the more adequate to describe bulk properties of
nuclear matter in the MFA.

\vspace{1.cm}

\begin{tabular}{cccccccc}
\hline \\
$\delta$&$g_{\sigma}^q$&$g_{\sigma}^B$&$g_\omega$&$M_N^*/M_N$&$
\kappa$&$B/B_0$&$R/R_0$  \\
	&	       &	      & 	 &	     &
 [MeV]	&	&	  \\ \hline
   3.6	&      2       &    6.30      &   10.04  &  0.70     &
 431	&  0.36 &  1.27   \\
   4	&	       &    6.13      &    9.65  &  0.72     &
 398	&  0.39 &  1.25   \\
   8	&	       &    5.65      &    8.54  &  0.77     &
 336	&  0.48 &  1.18   \\
  12	&	       &    5.53      &    8.29  &  0.76     &
 324	&  0.50 &  1.17   \\ \hline
 set II &    2.61      &	      &    6.67  &  0.78     &
 224	&  0.54 &  1.12   \\ \hline

\end{tabular}

\section{Aknowledgements}
This work was partially supported by the Universidad Nacional
de La Plata (UNLP) and the Consejo Nacional de Investigaciones
Cient\'{\i}ficas y T\'ecnicas (CONICET). I would like to thank to
Dr. R. Aguirre for his help and useful discussions. I thank to
Professor T. Kodama and the Centro Brasileiro de Pesquisas Fisicas
(CBPF) by their hospitality.

\end{document}